\documentclass[12pt]{article}
\usepackage{amsmath}
\usepackage{makeidx}
\usepackage{amsfonts}
\usepackage{amssymb}
\usepackage{graphicx}
\begin{document}

\title{A new class of binning-free, multivariate goodness-of-fit tests: the energy
tests\footnote{Supperted by Bunderministerium f\"{u}r Bildung und Forschung, Deutschland}}
\author{B. Aslan and G. Zech\thanks{Corresponding author, email: zech@physik.uni-siegen.de}\\Universit\"{a}t Siegen, D-57068 Siegen}
\maketitle
\begin{abstract}
We present a new class of multivariate binning-free and nonparametric
goodness-of-fit tests. The test quantity \emph{energy} is a function of the
distances of observed and simulated observations in the variate space. The
simulation follows the probability distribution function $f_{0}$ of the null
hypothesis. The distances are weighted with a weighting function which can be
adjusted to the variations of $f_{0}$. We have investigated the power of the
test for a uniform and a Gaussian distribution of one or two variates,
respectively and compared it to that of conventional tests. The energy test
with a Gaussian weighting function is closely related to the Pearson $\chi
^{2}$ test but is more powerful in most applications and avoids arbitrary bin
boundaries. The test is especially powerful in the multivariate case.
\end{abstract}

\section{Introduction}

There exist a large number of binning-free goodness-of-fit tests for
univariate distributions. The extension to the multivariate case is difficult,
because there a natural ordering system is missing. The popular $\chi^{2}$
test on the other hand suffers from the arbitrariness of the binning and from
lack of power for small samples.

The test proposed in this article avoids both, ordering and binning. It is
especially powerful in the multivariate case. It is called energy test,
because the definition of the test statistic is closely related to the energy
of electric charge distributions. To facilitate the computation of the test
statistic, we approximate the probability distribution function (p.d.f.)
$f_{0}$ of the null hypothesis by simulated observations. The statistic
``energy'' can also be used to test whether two samples belong to the same
parent distribution.

In Section 2 we define the test and study its properties. Section 3 contains a
comparison of the power of the energy test with that of other popular tests in
one dimension. In Section 4 we apply the energy test to two-dimensional
problems. We discuss possible modifications and extensions of the test in
Section 5 and conclude with a summary in Section 6.

\section{The test statistic}

\subsection{The energy function}

We define a quantity $\phi$, the \emph{energy}, which measures the difference
between two p.d.f's $f_{0}(\mathbf{x})$ and $f(\mathbf{x})$, $\mathbf{x}$
$\in\mathbb{R}^{n}$, by%

\begin{equation}
\phi=\frac{1}{2}\int\int\left[  f(\mathbf{x})-f_{0}(\mathbf{x})\right]
\left[  f(\mathbf{x}^{\prime})-f_{0}(\mathbf{x}^{\prime})\right]
R(\mathbf{x},\mathbf{x}^{\prime})\mathbf{dxdx}^{\prime}.\label{edef}%
\end{equation}

The integrals extend over the full variate space. The weight function $R$ is a
monotonically decreasing function of the Euklidian distance $|\mathbf{x}%
-\mathbf{x}^{\prime}|$. Relation (\ref{edef}) with $R=1/|\mathbf{x}%
-\mathbf{x}^{\prime}|$ is the electrostatic energy of two charge distributions
of opposite sign which is minimum if the charges neutralize each other.
Setting the function $R=\delta(\mathbf{x}-\mathbf{x}^{\prime})$, where
$\delta(\mathbf{x-x}^{\prime})$ is the Dirac Delta function, $\phi$ reduces to
the integrated quadratic difference of the two p.d.f.s
\[
\phi_{\delta}=\frac{1}{2}\int\left[  f(\mathbf{x})-f_{0}(\mathbf{x})\right]
^{2}\mathbf{dx.}%
\]

Since we want to generalize (\ref{edef}) in such a way that we can apply it to
a comparison of a sample with a distribution or of two samples with each
other, $\phi_{\delta}$ is not suitable. Functions like Gaussians which
correlate different locations have to be used.

The test which we will introduce is based on the fact that for fixed
$f_{0}(\mathbf{x})$, $\phi$ is minimum for the null hypothesis $f(\mathbf{x}%
)\equiv f_{0}(\mathbf{x})$ for all $\mathbf{x}$ $\in\mathbb{R}^{n}$. We sketch
a prove of this assertion in the Appendix.

Expanding (\ref{edef})
\[
\phi=\frac{1}{2}\int\int[f(\mathbf{x})f(\mathbf{x}^{\prime})+f_{0}%
(\mathbf{x})f_{0}(\mathbf{x}^{\prime})-2f(\mathbf{x})f_{0}(\mathbf{x}^{\prime
})R(|\mathbf{x}-\mathbf{x}^{\prime}|)]\mathbf{dxdx}^{\prime},
\]
we obtain three terms which have the form of expectation values of $R$. They
can be estimated from two samples $\mathbf{X}_{1},\mathbf{X}_{2}%
,...,\mathbf{X}_{N}$ and $\mathbf{Y}_{1},\mathbf{Y}_{2},...,\mathbf{Y}_{M}$
drawn from $f$ and $f_{0}$, respectively. Since in the first and the second
term a product of identical distributions occurs, there it is not necessary to
draw two different samples of the same p.d.f.. The expectation value of $R$ in
these expressions is given by the mean value of $R$ computed from all
combinations of the sample observations%

\[
\phi_{NM}=\frac{1}{N(N-1)}\sum_{j>i}R(|\mathbf{x}_{i}-\mathbf{x}_{j}%
|)+\frac{1}{M(M-1)}\sum_{j>i}R(|\mathbf{y}_{i}-\mathbf{y}_{j}|)-\frac{1}%
{NM}\sum_{i,j}R(|\mathbf{x}_{i}-\mathbf{y}_{j}|).
\]

Throughout this article unless specified differently, all sums run from $1$ to
the maximum value of the index. For the energy statistic $\phi_{N}$ of a
sample $\mathbf{X}_{1},\mathbf{X}_{2},...,\mathbf{X}_{N}$ relative to the
p.d.f. $f_{0}$ is given by:
\begin{align*}
\phi_{N}  & =\frac{1}{N(N-1)}\sum_{j>i}R(|\mathbf{x}_{i}-\mathbf{x}_{j}%
|)+\int\int f_{0}(\mathbf{x})f_{0}(\mathbf{x}^{\prime})R(|\mathbf{x}%
-\mathbf{x}^{\prime}|)\mathbf{dxdx}^{\prime}+\\
& -\frac{1}{N}\sum_{i}\int f_{0}(\mathbf{x}^{\prime})R(|\mathbf{x}%
_{i}-\mathbf{x}^{\prime}|)\mathbf{dx}^{\prime}.
\end{align*}

To test whether a statistical sample $\mathbf{X}_{1},\mathbf{X}_{2}%
,...,\mathbf{X}_{N}$ of size $N$ is compatible with the null hypothesis
$H_{0}$, we could in principle use the test statistic $\phi_{N}$ of the sample
with respect to the associated p.d.f. $f_{0}$ according to $\phi_{N}$ but
since the evaluation of $\phi_{N}$ usually requires a sum over difficult
integrals, we prefer to represent $f_{0}$ by a sample $\mathbf{Y}%
_{1},\mathbf{Y}_{2},...,\mathbf{Y}_{M}$, usually generated through a Monte
Carlo simulation. We drop in $\phi_{NM}$ the term depending only on $f_{0}$
which is independent from the sample observations and obtain:
\begin{equation}
\phi_{NM}=\frac{1}{N^{2}}\sum_{j>i}R(|\mathbf{x}_{i}-\mathbf{x}_{j}|)-\frac
{1}{NM}\sum_{i,j}R(|\mathbf{x}_{i}-\mathbf{y}_{j}|)\label{edefddm}%
\end{equation}

Furthermore, we have replaced the denominator of the first term $1/N(N-1)$ by
$1/N^{2}$ which has superior small sample properties (see Appendix).
Statistical fluctuations of the simulation are negligible if $M$ is large
compared to $N$, typically $M\geq10N$.

\subsection{The weight function}

We have investigated three different types of \ weight functions, power laws,
a logarithmic dependence and Gaussians.
\begin{align}
R_{pow}(r) &  =\left\{
\begin{array}
[c]{l}%
\frac{1}{r^{\kappa}}\;\text{for }r>d_{\min}\\
\frac{1}{d_{\min}^{\kappa}}\;\text{for }r\leq d_{\min}%
\end{array}
\right. \label{rpot1}\\
R_{\log}(r) &  =\left\{
\begin{array}
[c]{l}%
-\ln r\;\text{for }r>d_{\min}\\
-\ln d_{\min}\;\text{for }r\leq d_{\min}%
\end{array}
\right. \\
R_{G}(r) &  =\exp\left(  -r^{2}/(2s^{2})\right) \label{rexp}%
\end{align}

The first type is motivated by the analogy to electrostatics, the second is
long range and the third emphasizes a limited range for the correlation
between different observations. The power $\kappa$ of the denominator in
(\ref{rpot1}) and the parameter $s$ in (\ref{rexp}) may be chosen differently
for different dimensions of the sample space and different applications. For
slowly varying $f_{0}$ a small value of $\kappa$ around $0.1$ is recommended.
For short range variations the test quantity with larger values around $0.3$
is more sensitive.

Also the logarithmic function $R_{\log}$ is well adapted to slowly varying
$f_{0}$. The corresponding test quantiles are invariant under the
transformation $r\rightarrow ar$. For more rapidly varying $f_{0}$, the
Gaussian weight function $R_{G}$ is recommended. It permits to adjust the
range of the smearing to the shape of $f_{0}$.

The inverse power law and the logarithm exhibit a pole at $r$ equal to zero
which could produce infinities in the double discrete version of the energy
function $\phi_{NM}$. Very small distances, however, should not be weighted
too strongly since deviations from $f_{0}$ with sharp peaks are not expected
and usually inhibited by the finite experimental resolution. We eliminate the
poles by introducing a lower cutoff $d_{\min}$ for the distances $r$.
Distances less than $d_{\min}$ are replaced by $d_{\min}$. The value of this
parameter is not critical at all, it should be of the order of the average
distance of the simulation points in the region where $f_{0}$ is maximum.

\subsection{The distribution of the test statistic}%

\begin{figure}
[h]
\begin{center}
\includegraphics[
height=10.1067cm,
width=15.1259cm
]%
{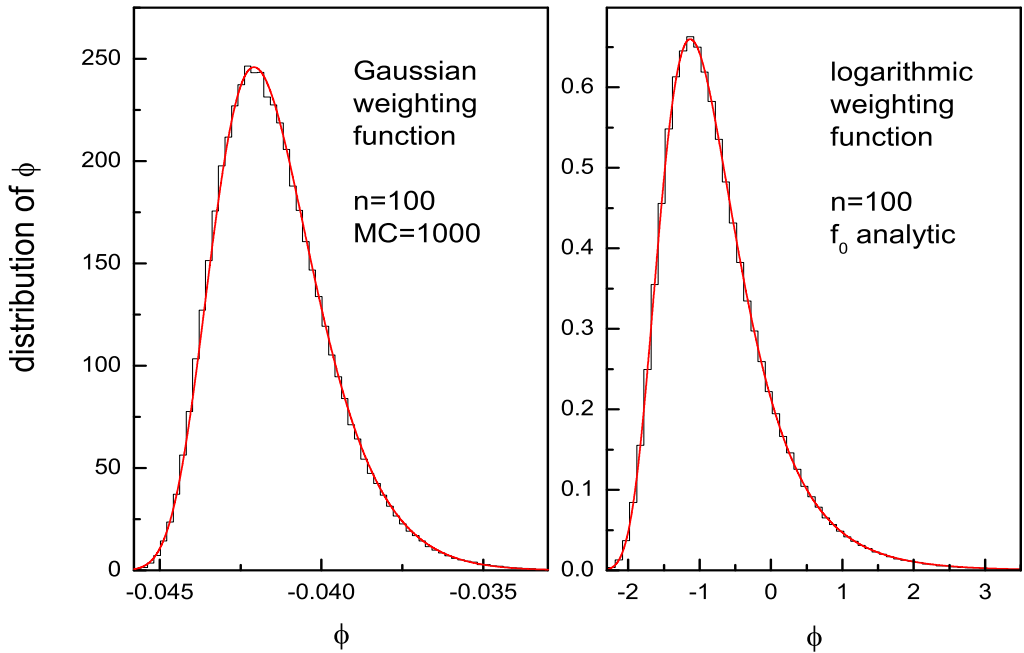}%
\caption{{\small Energy distributions for a Gaussian weighting function (left)
and a logarithmic weighting function (right) and their approximation by
extended extreme value distributions.}}%
\label{phidist}%
\end{center}
\end{figure}

The distribution of $\phi$ depends on the distribution function $f_{0}$ and on
the weight function $R$. Figure \ref{phidist} shows the distributions for
uniform $f_{0}$ with Gaussian and logarithmic weight functions. The
distributions are well described by a generalized extreme value distribution
\[
f(x)=\frac{1}{\sigma}\left(  1+\xi\frac{x-\mu}{\sigma}\right)  ^{-1/\xi-1}%
\exp\left\{  -\left(  1+\xi\frac{x-\mu}{\sigma}\right)  ^{-1/\xi}\right\}
\]
depending on three parameters, a scale parameter $\sigma$, a location
parameter $\mu$, and a shape parameter $\xi$. Rather than computing these
parameters from the moments of the specific $\phi$ distributions, we propose
to generate the distribution of the test statistic and the quantiles by a
Monte Carlo simulation. As a consequence of the dramatic increase of computing
power during the last decade, it has become possible to perform the
calculations on a simple PC within minutes. There is no need to publish tables
of percentage points, to distinguish between simple and composite hypotheses
and also censoring can be incorporated into the simulation.

\subsection{Relation to the Pearson $\chi^{2}$ test}

Let us assume that we have in a certain $\chi^{2}$-bin $n_{i}$ experimental
observations and $m_{i}$ Monte Carlo observations and $\beta=M/N$ $\gg1$. Then
the $\chi^{2}$ contribution of that bin is
\begin{align*}
\chi_{i}^{2}  & =\frac{(n_{i}-m_{i}/\beta)^{2}}{m_{i}/\beta}\\
& =\frac{n_{i}^{2}-2n_{i}m_{i}/\beta+m_{i}^{2}/\beta^{2}}{m_{i}/\beta}%
\end{align*}
For a goodness-of-fit test, an additive constant is irrelevant. Thus we can
drop $m_{i}^{2}$ in the last relation. If in addition the theoretical
distribution is uniform and the bins have constant size, we can also ignore
the denominator.
\[
\chi_{i}^{2\prime}=\frac{n_{i}^{2}}{2}-\frac{n_{i}m_{i}}{\beta}%
\]
Up to a constant factor this last relation corresponds nearly to the energy
defined in (\ref{edefddm}) for a weight function which is constant inside the
bin and zero outside:
\[
\phi_{NM}=\frac{n_{i}(n_{i}-1)}{2N^{2}}-\frac{n_{i}m_{i}}{NM}\approx\frac
{1}{N^{2}}\chi_{i}^{2\prime}%
\]

The $\chi^{2}$ test applied to a uniform distribution $f_{0}$ is equivalent to
an energy test with a box shaped weight function and fixed box locations.

Replacing the box function by the Gaussian weight function, the sharp cut at
the bin boundaries and the arbitrary location of the boundary for fixed
binning are avoided but the idea underlying the $\chi^{2}$ test is retained.

\subsection{Optimization of the test parameters in a simple case}

To study the dependence of the power of the test on the choice of the weight
function and its parameters, we have chosen for $H_{0}$ a uniform, univariate
p.d.f.$\ f_{0}$ restricted to the unit interval $[0,1]$. We determined the
rejection power with respect to contaminations of $f_{0}$ with a linear and
two different Gaussian distributions which represent a wide and a more local
distortion.
\begin{align}
f_{0}  & =1\nonumber\\
f_{1}(X)  & =2X\label{blin}\\
f_{2}(X)  & =c_{2}\exp\left(  -64(X-0.5)^{2}\right) \label{bgauss1}\\
f_{3}(X)  & =c_{3}\exp\left(  -256(X-0.5)^{2}\right) \label{bgauss2}%
\end{align}
We required 5\% significance level and computed the rejection power which is
equal to one minus the probability for an error of the second kind.

As a reference, we also computed the power of a $\chi^{2}$ test with bins of
fixed width. The number of bins $B\approx2N^{2/5}$ was chosen according to the
prescription proposed in \cite{moor86}.

The cut-off parameter $d_{\min}$ for $R_{\kappa}$ (power law) and $R_{l}$
(logarithmic) was set equal to $1/(4N)$.%

\begin{figure}
[ptb]
\begin{center}
\includegraphics[
height=20.1452cm,
width=15.1281cm
]%
{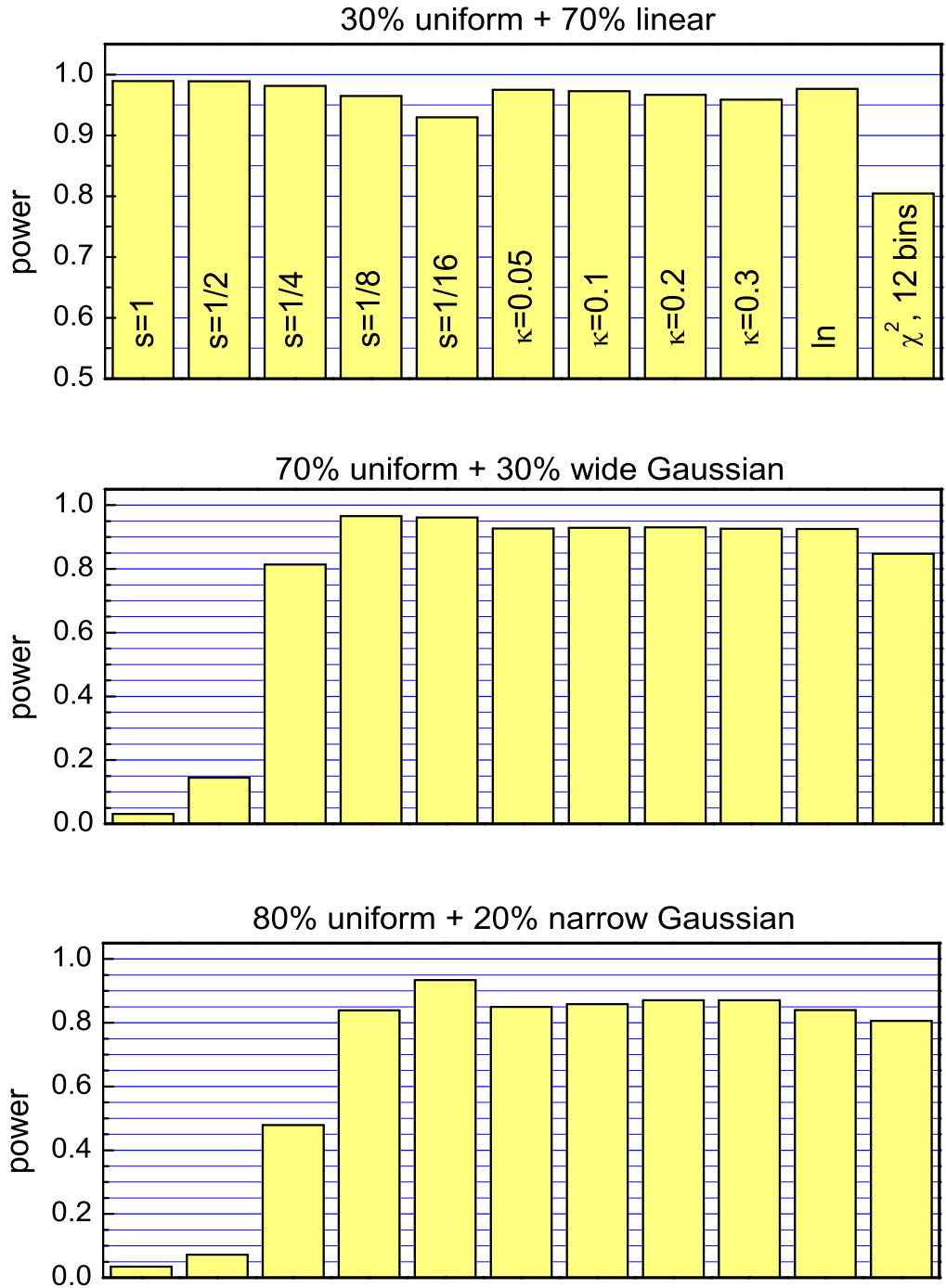}%
\caption{{\small Rejection power of different energy tests for a uniform
distribution }$f_{0}$ {\small with respect to a linear distribution and two
Gaussian contaminations, }$exp[-64(x-0.5)^{2}]${\small \ and }%
$exp[-256(x-0.5)^{2}]${\small .}}%
\label{e100all}%
\end{center}
\end{figure}

In Figure \ref{e100all} we show the results for samples of 100 observations
for 70 \% contamination with $f_{1}$, 30 \% contamination with $f_{2},$ and 20
\% contamination with $f_{3}$. Five different values of the Gaussian width
parameter $s$, four different power laws and the logarithmic weight function
have been studied.

As expected, the linear distribution is best discriminated by slowly varying
weight functions like the logarithm, low power laws and the wide Gaussians.

The contamination with the narrow Gaussian $f_{3}$ is better recognized by the
narrow weight functions ($s=1/8,\kappa=0.3$). Here the two wide Gaussian
weight functions fail completely.%

\begin{figure}
[ptb]
\begin{center}
\includegraphics[
height=20.1452cm,
width=15.1281cm
]%
{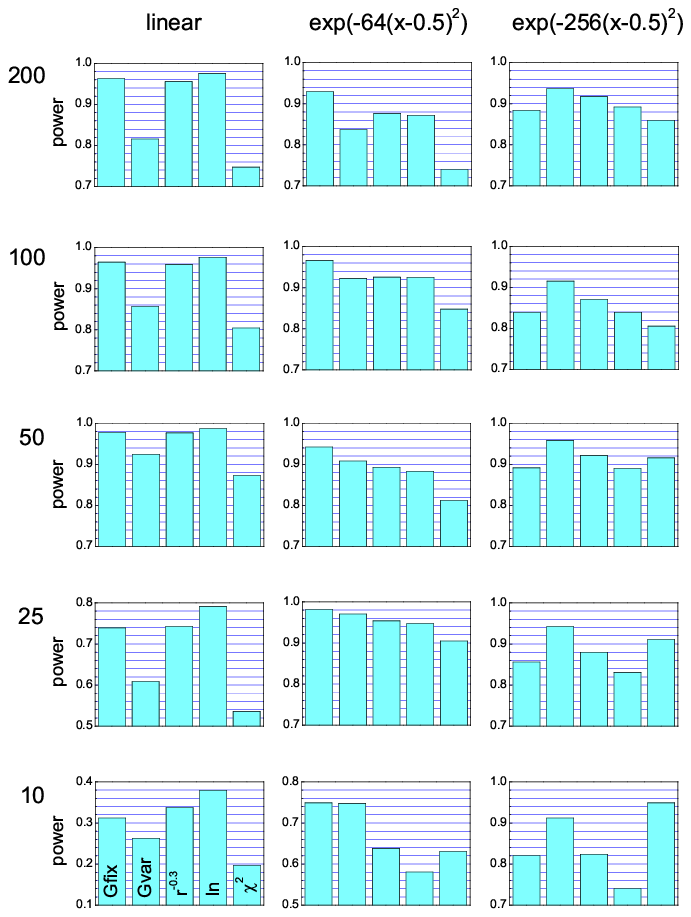}%
\caption{{\small Power of tests for 3 different contaminations to uniform
distribution and 5 different sample sizes ranging from 10 to 200. The shape of
the contamination is displayed on top of the columns. The type of test is
indicated in the lowest left hand plot.} }%
\label{power1}%
\end{center}
\end{figure}

Figure \ref{power1} illustrates the dependence of the rejection power on the
sample size for the three different distortions (\ref{blin}), (\ref{bgauss1}),
(\ref{bgauss2}) of the uniform distribution. The amount of contamination was
reduced with increasing sample size.

We applied the energy tests with power law $\kappa=0.3$, the logarithmic and
two Gaussian weight functions with fixed width $s=1/8$ (Gfix) and variable
width (Gvar). The variable width was chosen such that the full width at half
maximum is equal to the $\chi^{2}$ bin width, chosen according to the
$2N^{2/5}$ law. This allows for a fair comparison between the two methods.

As expected the linear distribution is best discriminated by the energy test
with logarithmic weight function. The power of the $\chi^{2}$ test is
considerably worse. The energy test with variable Gaussian weight function
follows the trend of the $\chi^{2}$ test but performs better in 14 out of the
15 cases.

Comparing the samples with sizes 50 and 100, respectively, we realize one of
the caveats of the $\chi^{2}$ test: For the sample of size 50 there are 9
bins. The central bin coincides favorably with the location of the distortion
peak of the background sample and consequently leads to a high rejection
power. For the sample of size 100, however, there are 12 bins, thus two bins
share the narrow peak and the power is reduced. The Gaussian energy test is
insensitive to the location of the distortion.

\section{Comparison with alternative univariate tests}

We have investigated the following goodness-of-fit tests, some of which have
been designed for special applications and can be found in \cite{dago86}:
$\chi^{2}$ test, Kolmogorov test, Kuiper test, Anderson-Darling test, Watson
test, Neyman smooth test, Region test, Energy tests with logarithmic and
Gaussian weight functions. The region test has been developed to detect
localized bumps by partitioning the variate range of the order statistic into
three regions with expected probabilities $p_{i},p_{j},1-p_{i}-p_{j}$. For
$N_{i}$ and $N_{j}$ observations observed in the first two regions, the test
statistic is
\[
R=\sup_{i,j}\{(N_{i}-Np_{i})^{2}+(N_{j}-Np_{j})^{2}+\left[  N_{i}%
+N_{j}-N(p_{i}+p_{j})\right]  ^{2}%
\]
where $N_{i}$ is the number of observations with $x\leq x_{i}$.

Samples contaminated by different background sources were tested against
$H_{0}$, corresponding to the uniform distribution. The power of each test for
a 5 \% significance limit was evaluated from $10^{5}$ Monte Carlo simulations
containing 100 observations each.

\begin{figure}
[ptb]
\begin{center}
\includegraphics[
trim=0.000000in 0.049849in 0.000000in 0.070153in,
height=6.0934cm,
width=15.1281cm
]%
{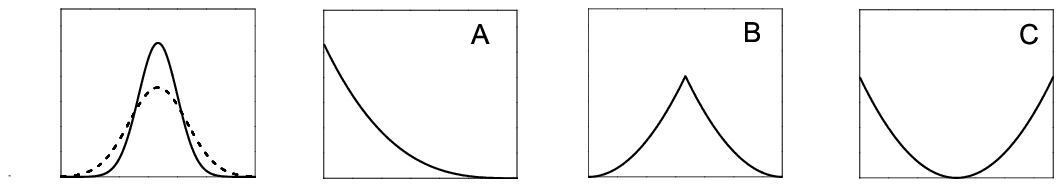}%
\caption{{\small Background distributions used to contaminate the uniform
distribution.}}%
\label{bgd1}%
\end{center}
\end{figure}

Samples with background contamination by a linear distribution, Gaussians and
the following distributions displayed in Figure \ref{bgd1} were generated by%

\begin{align*}
b_{A}(x)  & \varpropto(1-x)^{3}\\
b_{B}(x)  & \varpropto\left\{
\begin{array}
[c]{c}%
x^{2}\\
(1-x)^{2}%
\end{array}%
\begin{array}
[c]{c}%
\text{for }0\leq x<0.5\\
\text{for }0.5\leq x<1
\end{array}
\right. \\
b_{C}(x)  & \varpropto(x-0.5)^{2}\text{ \ \ \ \ \ \ \ }%
\end{align*}
.%

\begin{figure}
[ptb]
\begin{center}
\includegraphics[
height=20.1452cm,
width=15.1281cm
]%
{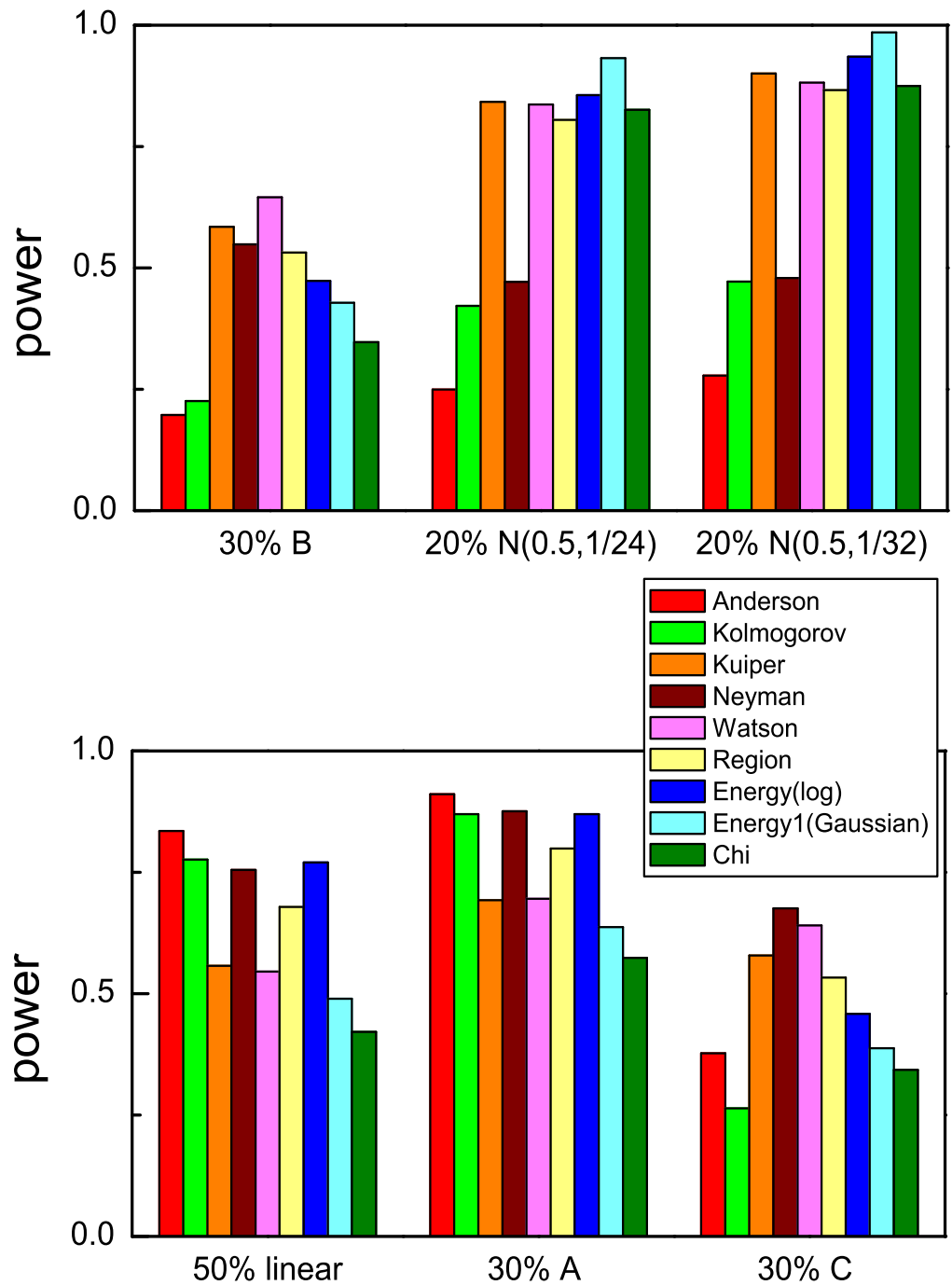}%
\caption{{\small Rejection power of tests with respect to different
contaminations to the uniform distribution. The contaminations correspond to
the distributions shown in the previous figure.} }%
\label{power1d}%
\end{center}
\end{figure}

The power of the different tests is presented in Figure \ref{power1d}. As
expected, none of the tests is optimum for all kind of distortions. The energy
tests are quite powerful independent of the background function.%

\section{The energy test for two or more variates}

While numerous goodness-of-fit tests exist for univariate distributions in
higher dimensions only tests of the $\chi^{2}$ type have become popular.

The extension of binning-free tests based on linear rank statistics to the
multivariate case \cite{just97} is difficult because no obvious ordering
scheme of the observations exist. Sequential ordering in the different
variates depends on the sequence in which the variates are chosen. In addition
to this logical difficulty there is a more fundamental problem: The
re-shuffling of the observations destroys the natural metric used to display
the data. The situation in goodness-of-fit problems is very similar to that in
pattern recognition. The possibility to detect distortions of the expected
distribution by background or resolution effects depends on an appropriate
selection of the random variables. The natural choice is usually given by the
experiment. By mapping a multivariate space onto a space defined by the
ordering recipe much of the information contained in the original distribution
may be lost.

\subsection{Test for bivariate Gaussian distribution}

It is rather common to deal with samples of observations which are drawn from
a multivariate normality. Multivariate goodness-of-fit tests have been
developed mostly for multinormality. An overview can be found in
\cite{dagos86}. An extension of the moment tests introduced by K. Pearson to
two variates is Mardia's test \cite{mard70}. Skewness $\beta_{1}$ and kurtosis
$\beta_{2}$ are compared to the nominal values for the normal distribution,
$\beta_{1}=0$ and $\beta_{2}=3$ respectively. We compared the logarithmic and
the Gaussian energy tests to Mardia's tests and to the two-dimensional Neyman
smooth test \cite{neymxx}.

We chose a standard Gaussian
\[
f_{0}=\frac{1}{2\pi}\exp\left(  -(x^{2}+y^{2})/2\right)
\]
to represent the null hypothesis.

\begin{figure}
[ptb]
\begin{center}
\includegraphics[
height=6.0934cm,
width=15.1545cm
]%
{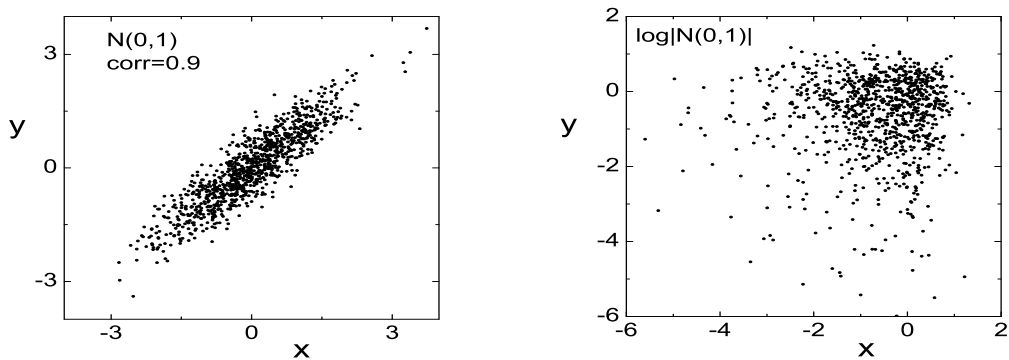}%
\caption{{\small Background distributions used to contaminate a
two-dimensional Gaussian.}}%
\label{bgd2}%
\end{center}
\end{figure}

Background samples added to $f_{0}$ are shown in Figure \ref{bgd2}. In
addition, a uniform background in the region ($0<X,Y<1$) and variations of the
Gaussian background were investigated. All Gaussians have the same width in
both coordinates $X$ and $Y$, but differ in the value of the width and the
correlation coefficient.%

\begin{figure}
[h]
\begin{center}
\includegraphics[
height=12.0441cm,
width=15.1281cm
]%
{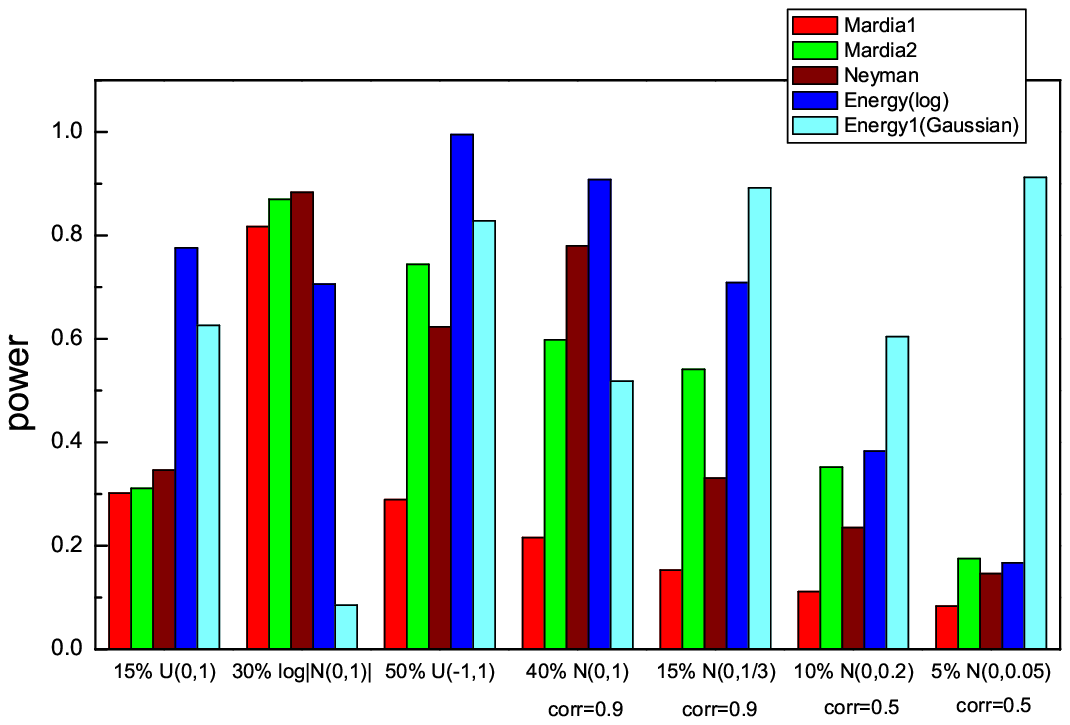}%
\caption{{\small Rejection power of tests with respect to different
contaminations of a two-dimensional Gaussian. The uniform background is
denoted by }$U${\small , the Gaussian contamination by }$N(mean,st.dev.)$.}%
\label{power2d}%
\end{center}
\end{figure}

Figure \ref{power2d} displays the results. We were astonished how well the
energy test competes with alternatives especially designed to test normality.
We attribute the excellent performance of the energy test to the fact that it
is sensitive to all deviations of two distributions whereas the Mardia tests
are based only on two specific moments.

\subsection{Example from particle physics}

\begin{figure}
[h]
\begin{center}
\includegraphics[
height=9.1028cm,
width=15.1281cm
]%
{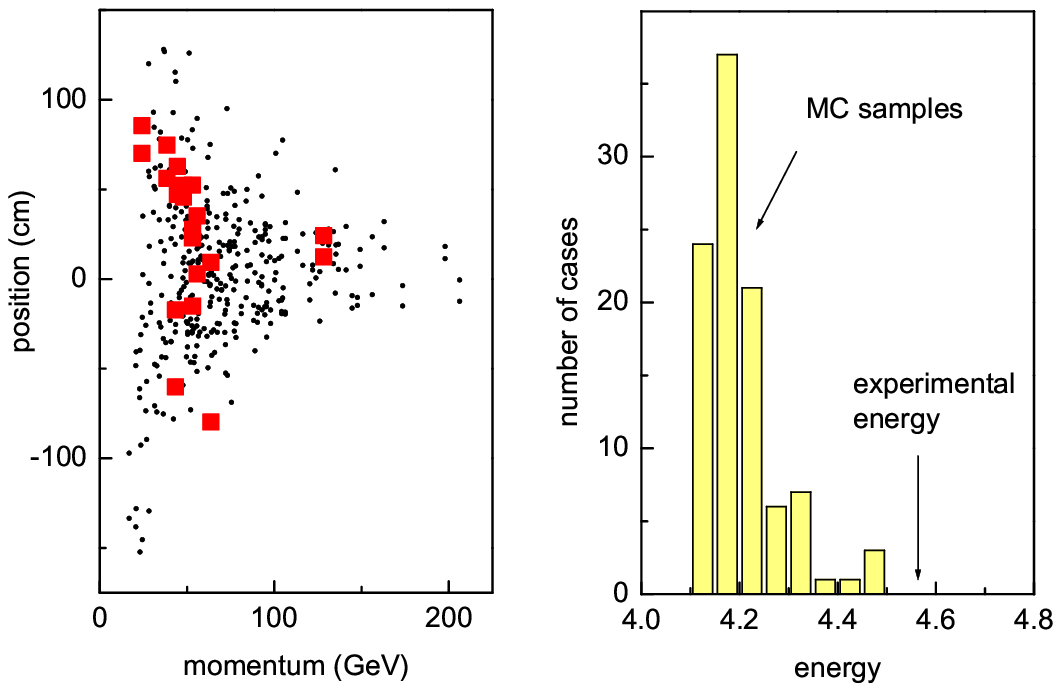}%
\caption{{\small Comparison of experimental distribution (squares) with Monte
Carlo simulation (black dots). The experimental energy computed from the
scatter plot (left) is compared to a Monte Carlo simulation of the experiment
(right).}}%
\label{scatter}%
\end{center}
\end{figure}

Figure \ref{scatter} is a scatter plot of data obtained in the particle
physics experiment HERA-B at DESY. The positions of tracks from $\psi$ decays
are plotted against the momentum of the $\psi$ meson. The 20 square entries
represent the experimental data, the black dots correspond to a Monte Carlo simulation.

We have computed the energy statistic $\phi$ with the logarithmic weighting
function for the experimental data relative to the Monte Carlo prediction.
Separately, the energy was determined for 100 independent Monte Carlo samples
of 20 events each to estimate the distribution of $\phi$ under $H_{0}$. The
experimental value is larger than all Monte Carlo values (see Figure
\ref{scatter}), indicating that the data do not follow the prediction.%

\section{Extensions and variations of the energy test}

\subsection{Normalization to probability density}

The Gaussian energy test has proven to be similar but superior to the
$\chi^{2}$ test for the uniform distribution. For non-uniform p.d.f.s the
similarity can be retained by effectively dividing the elements of the sums in
(\ref{edefddm}) by the p.d.f. $f_{0}$.
\[
R(|\mathbf{x}_{i}-\mathbf{x}_{j}|)\rightarrow R^{\prime}(|\mathbf{x}%
_{i}-\mathbf{x}_{j}|)=\frac{R(|\mathbf{x}_{i}-\mathbf{x}_{j}|)}{\sqrt
{f_{0}(\mathbf{x}_{i})f_{0}(\mathbf{x}_{j})}}%
\]
If the density is only given by a Monte Carlo simulation, the density can be
estimated from the volume of a multidimensional sphere in the variate space
containing a fixed number of Monte Carlo observations. In the example of
Figure \ref{scatter} the density at $\mathbf{x}$ could be estimated from the
area $A(\mathbf{x})$ of the smallest circle centered at $\mathbf{x}$
containing 10 Monte Carlo points: $f_{0}(\mathbf{x})\approx10/A(\mathbf{x})$.

\subsection{Width of the weighting function $R$ and metric}

Some statisticians recommend to use equal probability bins for the $\chi^{2} $
test in one dimension. An equivalent procedure for the Gaussian energy test
would be to adjust the width of the Gaussian weighting function to the p.d.f.
\begin{align*}
R_{s}(r_{ij})  & \rightarrow R_{s}^{\prime}(r_{ij})=\exp\left(  -r^{2}%
/(2s_{i}s_{j})\right) \\
s_{i}  & \varpropto1/f_{0}(\mathbf{x}_{i})
\end{align*}

Such an adjustment can be applied independent of the dimension of the variate
space while an equal probability binning is rather difficult in two or higher dimensions.

The resolution of $R$ should be adjusted to the expected deviations. The
constants $s$ can be chosen differently for the different coordinates.
Alternatively, a variate transformation may be performed before the test is applied.

The multi-dimensional logarithmic energy test requires a sensible choice of
the metric. The distances of the observations should be about equal
numerically in all directions of the variate space. This can be achieved by
normalizing the variates to the square root of their variance.

\subsection{Clustering of observations}

For very large samples, the computation of $\phi$ may become excessively long
even with powerful computers. A possible solution to this problem is to
combine observations to clusters. The cluster weight is equal to the sum of
the observations included in the cluster and its location is their center of
gravity. The details of the clustering are not important. The simplest method
consists in combining points in bins formed by a simple grid. Another
possibility could be the following: Observations and clusters (once clusters
have been formed) are chosen randomly and combined with all points and
clusters which are within a fixed maximum distance. The process is terminated
when the number of clusters is below an acceptable limit.

\subsection{Are two samples drawn from the same distribution?}

A frequently occurring problem is to check whether two random samples belong
to the same unknown parent distribution. In the framework of the ``energy''
concept we can use bootstrap or permutation methods to deduce a reference
energy distribution. The energy computed according to (\ref{edefddm}) is then
compared to this distribution. Quantiles can be used to measure the
compatibility of the samples.

\section{Summary and conclusions}

Energy tests represent a powerful alternative to conventional tests. To our
knowledge the energy test is the first multi-variate binning-free test which
is independent of a subjective ordering system. With a logarithmic weighting
function it is very sensitive to long range distortions of the distribution to
be tested, a situation which is frequent in physics applications. With a
Gaussian weighting function the energy test has similar properties as tests of
the $\chi^{2}$ type, but avoids arbitrary bin boundaries. It is more powerful
than the $\chi^{2}$ test in almost all cases which we have considered. It can
cope with arbitrarily small sample sizes. It is astonishing how well the
energy test can compete with the Mardia test for testing normality in two dimensions.

The necessity to determine the distribution function of the test statistic for
the specific application is not a serious restriction with today's computing power.

The energy statistic can be used to test whether different samples belong to
the same parent distribution. A corresponding study will be presented in a
forthcoming article.

\section{Appendix}

\subsection{Continuous case}

We conjecture that $f(\mathbf{x})\equiv f_{0}(\mathbf{x})$ corresponds to a
minimum of $\phi$ in (\ref{edef}). Substituting $g(\mathbf{x})=f(\mathbf{x}%
)-f_{0}(\mathbf{x})$ we obtain
\begin{equation}
\phi=\frac{1}{2}\int\int g(\mathbf{x})g(\mathbf{x}^{\prime})R(|\mathbf{x}%
-\mathbf{x}^{\prime}|)\mathbf{dxdx}^{\prime}\geq0\label{phigg}%
\end{equation}
where $g$ fulfils the condition $\int g(\mathbf{x})\mathbf{dx}=0$. We
conjecture that $\phi$ is minimum for $g(\mathbf{x})\equiv0$ for weight
functions $R$ which decrease continuously with $|\mathbf{x}-\mathbf{x}%
^{\prime}|$.

In the following, we sketch a prove of this conjecture without full
mathematical rigor and generality.

The property (\ref{phigg}) is invariant under a linear transformation
\[
R\rightarrow R^{\prime}=\alpha R+const
\]
where $\alpha$ is a positive scaling factor. Assuming that $R$ is finite, we
can restrict its values to the range $-1\leq R\leq1$ with $R(0)=1$.\ We
approximate the function $g$ by a histogram with function value $g_{i}$ in bin
$i$. The size of each bin is $\Delta^{n}$.
\[
\phi_{h}=\Delta^{2n}\sum_{i,j}g_{i}g_{j}R_{ij}%
\]
with self-explaining notation. The weights can be substituted by cosines,
$\cos\theta_{ij}=R_{ij}$ with $\cos\theta_{ii}=1$ in our approximation. The
sum then can be written as a sum of vectors $\mathbf{g}_{i}$ in $\mathbb{R}%
^{n}$:
\begin{align*}
\phi_{h}  & =\Delta^{2n}\sum_{i,j}g_{i}g_{j}\cos\theta_{ij}\\
\phi_{h}  & =\Delta^{2n}\left(  \sum_{i}\mathbf{g}_{i}\right)  ^{2}\geq0
\end{align*}
The minimum of $\phi_{h}$ is realized only if all $\mathbf{g}_{i}$ are equal
to zero.

\subsection{Discrete case}

When we approximate both densities $f$ and $f_{0}$ by distributions of $N$
points each, with weights equal $1/N$, we require that the energy be minimum
if the two samples coincide, $\mathbf{x}_{j}=\mathbf{y}_{j}$ $\forall j$. To
fulfil this condition, we have to replace the factors $1/(N-1)$ and $1/(M-1)$
by $1/N$ in $\phi_{NM}$.
\begin{align*}
\phi_{NN}  & =\frac{1}{N^{2}}\sum_{j>i}R(|\mathbf{x}_{i}-\mathbf{x}%
_{j}|)+\frac{1}{N^{2}}\sum_{j>i}R(|\mathbf{y}_{i}-\mathbf{y}_{j}|)+\\
& -\frac{1}{N^{2}}\sum_{i,j}R(|\mathbf{x}_{i}-\mathbf{y}_{j}|)
\end{align*}
To demonstrate the minimum condition which leads to $\phi_{NN}=0$, we apply an
infinitesimal shift to one observation. For $\mathbf{x}_{j}=\mathbf{y}_{j} $
for $j\neq k$ and $\mathbf{x}_{k}-\mathbf{y}_{k}=\delta\mathbf{x}_{k} $, only
the pair $\mathbf{x}_{k},\mathbf{y}_{k}$ contributes to the energy, all other
terms cancel.
\[
\phi_{NN}=-\frac{1}{N^{2}}\left[  R(|\delta\mathbf{x}_{k}|-R(0)\right]
\]
Since $R$ \emph{decreases} with its argument, $R(0)-R(|\delta\mathbf{x}%
_{k}|)>0$, we have found a local \emph{minimum} of the energy.

For the special choice where $R$ is the distance function, i. e.
\emph{increasing} with the distance, it has been demonstrated by \cite{morg01}
that $\phi$
\[
\phi=\frac{1}{N^{2}}\sum_{j>i}\left[  |\mathbf{x}_{i}-\mathbf{x}%
_{j}|+|\mathbf{y}_{i}-\mathbf{y}_{j}|\right]  -\frac{1}{N^{2}}\sum_{i}\sum
_{j}|\mathbf{x}_{i}-\mathbf{y}_{j}|\leq0
\]
is \emph{maximum} for $\mathbf{x}_{j}=\mathbf{y}_{j}$ for all $j$. Since
$\phi$ is constant for constant $R$, it is plausible that for discrete
distributions $\phi$ is minimum for $\mathbf{x}_{j}=\mathbf{y}_{j}$ for all
decreasing functions of the distance, and maximum for all increasing functions
of the distance, however, we did not succeed in finding a general prove of
this conjecture.

\bigskip

\textbf{Acknowledgments. }The authors wish to express their gratitude to Prof.
R.-D. Reiss for his constructive comments on earlier drafts of this paper,
which led to improvements in the presentation.

\end{document}